\documentclass[shortAfour,sageh,doublespace]{sagej}

\usepackage{lineno,hyperref}

\usepackage{todonotes}
\usepackage{makeidx}
\usepackage{graphics}
\usepackage{color}
\usepackage{listings}
\usepackage{bm}
\usepackage{url}
\usepackage{booktabs}
\usepackage{amsmath}
\usepackage{paralist}
\usepackage{amssymb}
\usepackage{amsthm}
\usepackage{subcaption}
\usepackage{float} % TODO: remove
\usepackage{adjustbox}

\newcommand{\bc}{\begin{center}}
\newcommand{\ec}{\end{center}}
\newcommand{\bd}{\begin{description}}
\newcommand{\ed}{\end{description}}
\newcommand{\bi}{\begin{itemize}}
\newcommand{\ei}{\end{itemize}}
\newcommand{\benu}{\begin{enumerate}}
\newcommand{\eenu}{\end{enumerate}}
\newcommand{\bq}{\begin{quote}}
\newcommand{\eq}{\end{quote}}
\newcommand{\be}{\begin{equation}}
\newcommand{\ee}{\end{equation}}
\newcommand{\bea}{\begin{eqnarray}}
\newcommand{\eea}{\end{eqnarray}}

\newcommand{\T}{T}
\newcommand{\dt}{{\Delta t}}

\newcommand{\pii}{{\partial_i}}
\newcommand{\pj}{{\partial_j}}

\definecolor{g90}{gray}{.90}
\definecolor{myBlue}{rgb}{0,0,.59}
\definecolor{myGreen}{rgb}{0,.20,.2}
\definecolor{myBrown}{rgb}{.59,0,0}

\begin{document}

%%%%%%%%%%%%%%%%%%%%%%%%%%%%%%%%%%%%%%%%%%%%%%%%%%%%%%%%%%%%%%%%%%%%%%%%

\runninghead{Calore, Gabbana, Schifano, Tripiccione}

%%%%%%%%%%%%%%%%%%%%%%%%%%%%%%%%%%%%%%%%%%%%%%%%%%%%%%%%%%%%%%%%%%%%%%%%

\title{Optimization of Lattice Boltzmann Simulations on Heterogeneous Computers}

%%%%%%%%%%%%%%%%%%%%%%%%%%%%%%%%%%%%%%%%%%%%%%%%%%%%%%%%%%%%%%%%%%%%%%%%

\author{
E. Calore\affilnum{1},
A. Gabbana\affilnum{1},
S.~F.~Schifano\affilnum{1},
R. Tripiccione\affilnum{1}
}
\affiliation{\affilnum{1}University of Ferrara and INFN Ferrara, 
via Saragat 1, I-44122 Ferrara, ITALY.}
\corrauth{Sebastiano Fabio Schifano, University of Ferrara and INFN Ferrara,
via Saragat 1, I-44122 Ferrara, ITALY.}
\email{schifano@fe.infn.it}
%
%%%%%%%%%%%%%%%%%%%%%%%%%%%%%%%%%%%%%%%%%%%%%%%%%%%%%%%%%%%%%%%%%%%%%%%%

\begin{abstract}
High-performance computing systems are more and more often based on 
accelerators.
Computing applications targeting those systems often follow a host-driven 
approach in which hosts offload almost all compute-intensive 
sections of the code onto accelerators; this approach only marginally 
exploits the computational resources available on the host CPUs, limiting 
performance and energy efficiency.
The obvious step forward is to run compute-intensive kernels in a 
concurrent and balanced way on both hosts and accelerators. 
In this paper we consider exactly this problem for a class of applications 
based on Lattice Boltzmann Methods, widely used in computational 
fluid-dynamics. 
Our goal is to develop just one program, portable and able to run efficiently 
on several different combinations of hosts and accelerators.  
To reach this goal, we define common data layouts enabling the code to exploit
efficiently the different parallel and vector options of the various
accelerators, and  matching the possibly different requirements of the
compute-bound and memory-bound  kernels of the application.
We also define models and metrics that predict the best partitioning of 
workloads among host and accelerator, and the optimally achievable overall 
performance level.
We test the performance of our codes and their scaling properties using as 
testbeds HPC clusters incorporating different accelerators: 
Intel Xeon-Phi many-core processors, NVIDIA GPUs and AMD GPUs.
\end{abstract}

 
%%%%%%%%%%%%%%%%%%%%%%%%%%%%%%%%%%%%%%%%%%%%%%%%%%%%%%%%%%%%%%%%%%%%%%%%

\maketitle

%%%%%%%%%%%%%%%%%%%%%%%%%%%%%%%%%%%%%%%%%%%%%%%%%%%%%%%%%%%%%%%%%%%%%%%%

\section{Background and related works}\label{sec:introduction}

The architecture of high-performance computing (HPC) systems is increasingly
based on accelerators; typical HPC systems today -- including several leading 
entries of the TOP500 list (\cite{top500}) -- are large clusters of processing 
nodes interconnected by a fast low-latency network, each node containing 
standard processors coupled with one or more accelerator units.\\
Accelerators promise higher computing performance and better energy efficiency,
improving on traditional processors up to one order of magnitude; in fact, they
i) allow massive parallel processing by a combination of a large number of processing 
cores and vector units, 
ii) allow for massive multi-threading, in order to hide memory access latencies and, 
iii) trade a streamlined control structure (e.g., executing instructions in-order only) 
for additional data-path processing for a given device or energy budget.
%Consequently, accelerators are at their best when handling lengthy, compute 
%intensive sections of application codes that offer a large amount of 
%exploitable parallelism.\\

Typical accelerated applications usually follow a host-driven 
approach in which the host processor offloads (almost) all compute-intensive 
sections of the code onto accelerators; the host itself typically only 
orchestrates global program flow or processes sequential segments of the 
application; this approach wastes a non negligible
amount of the available computational resources, reducing  
overall performance and energy efficiency.

The obvious step forward is that even compute-intensive application kernels
should be executed in a balanced way on both hosts and accelerators. 
This improvement has been hampered so far by several non trivial obstacles, especially 
because CPUs and accelerators often present different architectures, 
so efficient accelerated codes may involve 
different data structures and operation schedules. 
Moreover, the lack of well established performance-portable programming heterogeneous frameworks 
has so far required the use of specific programming languages (or at least proprietary 
variants of standard languages) on each different accelerator, harming portability, 
maintainability and possibly even correctness of the application.

Improvements on this latter aspect come with the recent evolution of  
{\em directive}-based programming environments, allowing programmers to annotate 
their codes with hints to the compiler about available parallelization options.
Several such frameworks have been proposed, such as the Hybrid 
Multi-core Parallel Programming model (HMPP) proposed by CAPS, hiCUDA (\cite{hicuda}), 
OpenMPC (\cite{openmpc}) and StarSs (\cite{starss}).
However, the most common compiler frameworks currently used for scientific codes are 
OpenMP (\cite{openmp}) and OpenACC (\cite{openacc}). Both frameworks allow to annotate codes written 
in standard languages (e.g., C, C+ and Fortran) with appropriate {\em pragma} directives 
characterizing the available parallelization space of each code section. 
This approach leaves to compilers to apply all optimization steps specific 
to each different target architecture and consistent with the directives, 
allowing in principle portability of codes between any supported host and 
accelerator device.
This process is still immature, and significant limits to portability still exist.
The OpenMP standard, version 4 (\cite{openmp4}), has introduced support for 
-- in principle -- any kind of accelerator, 
but compilers supporting GPUs are not yet available, and the Intel Xeon Phi 
is de-facto the only supported accelerator.
On the other hand OpenACC supports several different GPUs and more recently also 
multi-core CPU architectures, but not the Xeon-Phi, and does not allow to compile 
codes able to spread parallel tasks concurrently on both GPU and CPU cores.
Also, both standards do not address processor-specific hardware features so 
non-portable proprietary directives and instructions are often necessary; for example performance 
optimization on Intel CPU processors often requires Intel-proprietary compiler directives.
 
In spite of these weaknesses, and in the hope that a converging trend is in progress, 
one would like to
i) understand how difficult it is to design one common code using common domain 
data structures running concurrently on host(s) and accelerator(s), 
that is portable and also performance-portable across traditional processors and different accelerators, 
and ii) quantify the performance gains made possible by concurrent execution 
on host and accelerator. 

In order to explore this problem, one has to understand the impact on
performance  that different data layouts and execution schedules have for
different accelerator  architectures.  
One can then define a common data layout and  write a common code for a given
application with optimal (or close to optimal)  performance on several
combinations of host processors and accelerators. 
Assuming we can identify a common data layout giving good  
performance on several combinations of host processors and accelerators, 
then one has to find efficient partitioning criteria to split the execution 
of the code among hosts and processors.

% Cosa abbiamo fatto in questo lavoro
In this paper we tackle exactly these issues for a class of applications based
on Lattice Boltzmann (LB) methods, widely used in computational fluid-dynamics.
This class of applications offers a large amount of easily identified available 
parallelism, making LB an ideal target for accelerator-based HPC systems.
We consider alternate data layouts, processing schedules and optimal ways 
to compute concurrently on host and accelerator. We quantify the impact on 
performance, and use these findings to develop production-grade massively 
parallel codes.
We run benchmarks and test our codes on HPC systems whose nodes have dual Intel
Xeon processors and a variety of different accelerators, namely the NVIDIA K80
(\cite{k80-whitepaper}) and AMD Hawaii GPUs (\cite{amd-hawaii-whitepaper}), 
and the Intel Xeon-Phi (\cite{knc-whitepaper}).

Over the years, LB codes have been written and optimized for large clusters 
of commodity CPUs (\cite{Pohl2004}) and for application-specific 
machines (\cite{cimento09,iccs10,ppam13}).
More recent work has focused on exploiting the parallelism of powerful 
traditional many-core processors (\cite{caf13}), and of power-efficient 
accelerators such as GP-GPU (\cite{lbm-gpu2,cafgpu13,parco16}) and Xeon-Phi 
processors (\cite{iccs13}), and even FPGAs (\cite{lbm-fpga}).

Recent analyses of optimal data layouts for LB have been 
made in~\cite{Wittmann,shet1,shet2}. 
However, \cite{Wittmann} focuses only on the {\em propagate} step, one of the 
two key kernels in LB codes, while \cite{shet1} does not take into account 
vectorization; in \cite{shet2} vectorization is considered using intrinsic
functions only. None of these papers considers accelerators.
In \cite{ppam15} we have started preliminary investigation considering only 
the Xeon-Phi as an accelerator.
Here, we extend these results in several ways: first, we take into account both 
{\em propagate} and {\em collision} steps used in LB simulations. Then we 
use a high level approach based on compiler directives, and we take into 
account also NVIDIA and AMD accelerators commonly used in HPC communities.
Very recently, \cite{pedro15,pedro16} have explored the benefits of LB solvers on 
heterogeneous systems considering different memory layouts and system based both
on NVIDIA GPUs and Xeon-Phi accelerators. In our contribution we consider a
more  complex LB solver (D2Q37 instead of D2Q9) a wider analysis of data
layouts and an automatic analytic way to find the optimal partitioning of 
lattice domains between host-CPU and accelerators. 

% STruttura del paperello
This paper is structured as follows: \autoref{sec:architectures} introduces 
the main hardware features of CPUs and GPUs that we have taken into account 
in this paper and  \autoref{sec:lbm} gives an overview of LB methods. 
Section \ref{sec:implementation} discusses the design and implementation options 
for data layouts suitable for LB codes and measures the impact of different 
choices in terms of performances, while \autoref{sec:heterogeneous} describes 
the implementation of codes that we have developed concurrently running 
on both host and accelerator.
Section \ref{sec:optimization} describes two important optimization steps for our 
heterogeneous code and defines a performance model, while
\autoref{sec:scalability} analyzes our performance results on two 
different clusters, one with K80 GPUs and one with Xeon-Phi accelerators.
Finally, \autoref{sec:conclusion} wraps up our main results and highlights our 
conclusions.

%%%%%%%%%%%%%%%%%%%%%%%%%%%%%%%%%%%%%%%%%%%%%%%%%%%%%%%%%%%%%%%%%%%%%%

\section{Architectures of HPC systems}\label{sec:architectures}

%Here we discuss hardware architectures of standard processors, 
%accelerators and currently available HPC systems.

Heterogeneous systems have recently enjoyed growing popularity in the HPC landscape. 
These systems combine within a single processing node commodity multi-core 
architectures with off-chip accelerators, 
GPUs, Xeon-Phi many-core units and (sometimes) FPGAs. 
This architectural choice comes from an attempt 
to boost overall performance adding an additional processor (the accelerator) 
that exploits massively parallel data-paths to increase performance 
(and energy efficiency) at the cost of reduced programming flexibility.  
In this section we briefly review the architectures of the state-of-the-art 
accelerators that we have considered -- GPUs and Xeon-Phi many-core processors -- 
focusing on the impact that their diverging architectures have on the possibility 
to develop a common HPC code able to run efficiently on both of them.

GPUs are multi-core processors with a large number of processing units, all
executing in parallel. The NVIDIA K80 is a dual GK210 GPU,
each containing 13 processing units, called streaming multiprocessors (SMX).
Each processing unit has 192 compute units, called CUDA-cores,
concurrently executing groups of 32 operations in a SIMT (Single Instruction,
Multiple Thread) fashion; much like traditional SIMD processors, cores within
a group execute the same instruction at the same time but are allowed to take
different branches (at a performance penalty). The AMD FirePro W9100 is conceptually similar
to the K80 NVIDIA GPU; it has 44 processing units, each one with 64
compute units (stream processors).

The clock of an NVIDIA K80 has a frequency of 573 MHz which can be boosted
up to 875 MHz. The aggregate peak performance is then of 5.6 TFlops in single
precision and 1.87 TFlops in double precision (only one third of the SMXs work
concurrently when performing double precision operations). Working at 930MHz the
processing units of the AMD FirePro W9100 delivers up to 5.2 TFlops in SP and
2.6 in DP.

%%%%%%%%%%%%%%%%%%%%%%%%%%%%%%%%%%%%%%%%%%%%%%%%%%%%%%%%%%%%%%%%%%%%%%
\begin{table}
\centering 
\begin{adjustbox}{width=0.49\textwidth}
\begin{tabular}{l r r r r r} 
\toprule 
                 &  Xeon E5-2630 v3 & Xeon-Phi 7120P & Tesla K80 & FirePro W9100 \\   
\midrule 
\#physical-cores &  8               & 61             & 2~$\times$~13 SMX  & 44        \\ 
\#logical-cores  &  16              & 244            & 2~$\times$~2496    & 2816      \\    
Clock (GHz)      &  2.4             & 1.238          & 0.560     & 0.930     \\ 
Peak perf. (DP/SP GF) &  307/614         & 1208/2416      & 1870/5600 & 2620/5240 \\  
SIMD unit        &  AVX2 256-bit    & AVX2 512-bit   & N/A       & N/A       \\ 
LL cache (MB)    &  20              & 30.5           & 1.68      & 1.00      \\ 
\#Mem. Channels  &  4               & 16             & --        & --        \\ 
Max Memory (GB)  &  768             & 16             & 2~$\times$~12      & 16        \\ 
Mem BW (GB/s)    &  59              & 352            & 2~$\times$~240     & 320       \\ 
%ECC              & YES              & YES            & YES       & YES       \\ 
\midrule
\end{tabular} 
\end{adjustbox}
\caption{
Selected hardware features of the systems tested in this work:
Xeon E5-2630 is a commodity processor adopting the Intel Haswell micro-architecture;
Xeon-Phi 7120P is based on the Intel MIC architecture;
Tesla K80 is a NVIDIA GPU with two Tesla GK210 accelerators;
FirePro W9100 is an AMD Hawaii GPU
}\label{tab:arch-data}
\end{table}
%%%%%%%%%%%%%%%%%%%%%%%%%%%%%%%%%%%%%%%%%%%%%%%%%%%%%%%%%%%%%%%%%%%%%%

In general, GPUs sustain their huge potential performance thanks to large memory
bandwidth --in order to avoid starving the processors -- and massive
multi-threading -- to hide memory access latency. Consequently, register files
are huge in GPUs, as they have to store the states of many different threads,
while data caches are less important. For example (see also
\autoref{tab:arch-data}), a K80 GPU has a combined peak memory bandwidth of 480
GB/s, while each SMX has a register file of 512KB and just 128KB L1 cache/shared
memory; SMX units share a 1536KB L2 cache. Similarly, the AMD W9100 has a peak
memory bandwidth of 320 GB/s and its last level cache is just 1MB.

%%%%%%%%%%%%%%%%%%%%%%%%%%%%%%%%%%%%%%%%%%%%%%%%%%%%%%%%%%%%%%%%%%%%%%
%
\begin{figure*}
\centering
\begin{minipage}{0.32\textwidth}
\includegraphics[width=0.9\textwidth]{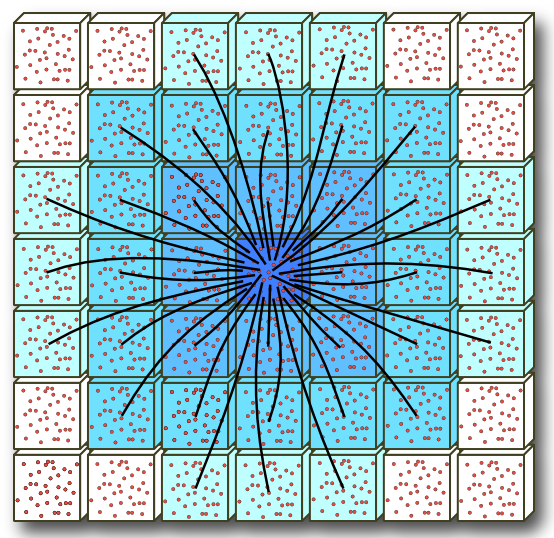}
\end{minipage}
\hspace*{18mm}
\begin{minipage}{0.32\textwidth}
\includegraphics[width=0.9\textwidth]{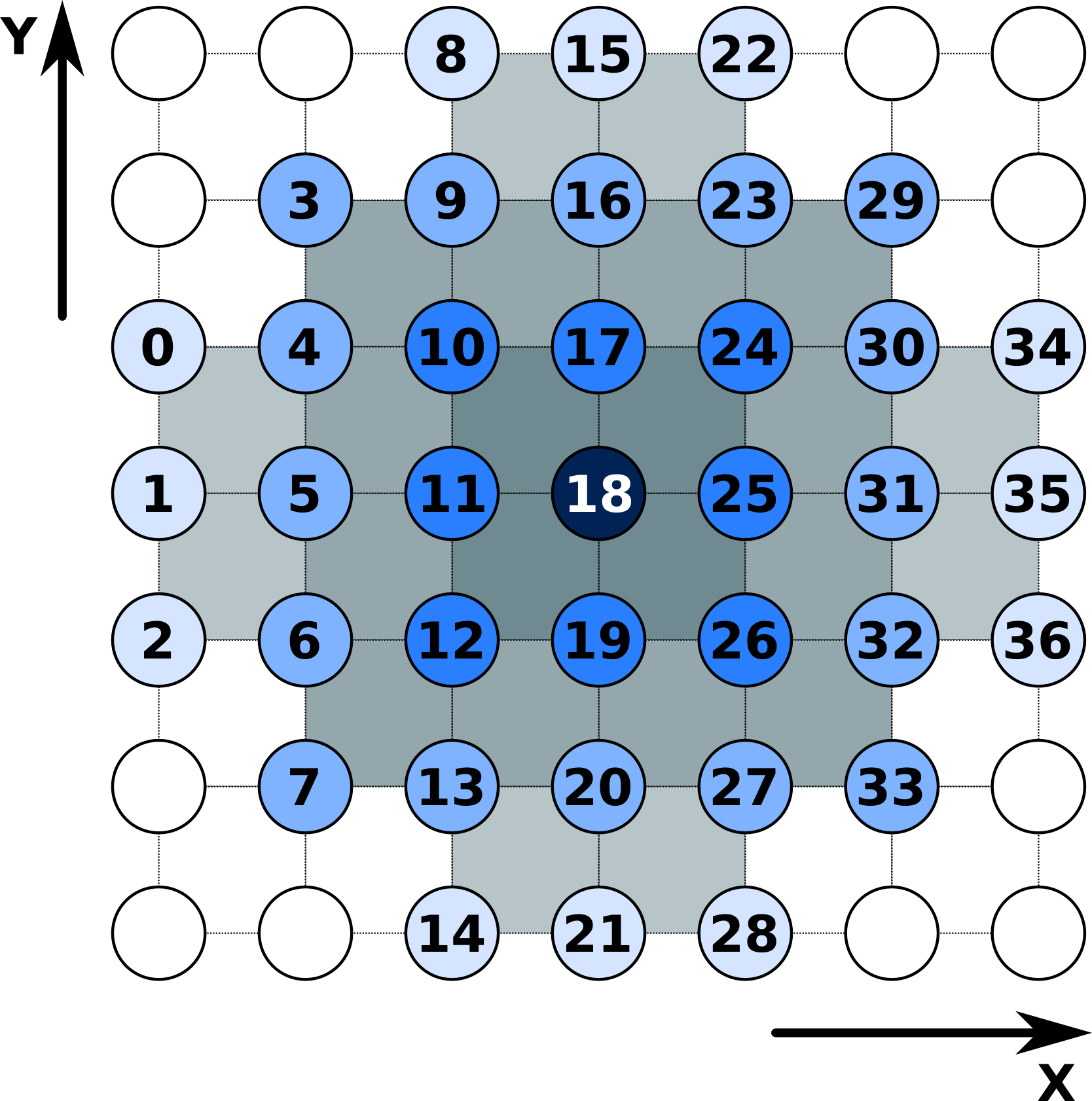}
\end{minipage}
\caption{Left: Velocity vectors for the LB populations of the D2Q37 model. 
Right: populations are identified by an arbitrary label, associated to the
lattice hop that they perform in the {\em propagate} phase.}
\label{fig:streamscheme}
\end{figure*}
%
%%%%%%%%%%%%%%%%%%%%%%%%%%%%%%%%%%%%%%%%%%%%%%%%%%%%%%%%%%%%%%%%%%%%%%

The architecture of Xeon-Phi processors -- the other class of accelerators 
that we consider -- builds on a very large
number of more traditional X86 cores, each optimized for streaming parallel
processing, with a streamlined and less versatile control part and enhanced
vector processing facilities. For instance, the currently available version of
this processor family -- the Knights Corner (KNC) --  integrates up to 61 CPU
cores, each supporting the execution of 4 threads, for an aggregate
peak performance of 1 TFlops in double precision, with a clock running at 1.2 GHz. Each
core has a 32 KB L1-cache and a 512 KB L2-cache. The L2-caches, private at the
core level, are interconnected  through a bi-directional ring and data is kept
coherent and indirectly accessible by all cores. The ring also connects to
a GDDR5 memory bank of 16 GB, with a peak bandwidth of 352 GB/s.
Each core has a Vector Processing Unit (VPU), executing SIMD operations on
vectors of 512 bits. 
 
Present generation GPUs and Xeon-Phi processors are connected to their host 
through a PCI-express interface, allowing data exchange between the two processors.
Typically 16 PCI-express lanes are used, with an aggregate bandwidth of 8 GB/s,
much smaller that the typical memory bandwidth of these processors, 
so processor-accelerator data exchanges may easily become serious performance bottlenecks. 

For both processor classes, in this work we consider a host-driven heterogeneous 
programming model, with applications executing both on the host and on the accelerator
\footnote{the Xeon-Phi also support ``native mode'' thus acting as an
independent node capable of running applications independently.  This approach
will be enhanced in the next generation Xeon-Phi system,  the Knights Landing,
that will also be available as a stand alone processor. We do not consider this
mode of operation in this paper.}.
So far, different programming languages have been available for each 
specific accelerator; indeed, NVIDIA GPUs have a proprietary 
programming language, and AMD GPUs are supported by the OpenCL 
programming environment.
On the contrary, the Xeon-Phi uses the same programming environment 
as its Xeon multi-core counterpart, so one can develop codes following 
a directive-based approach (e.g. OpenMP), slightly reducing the effort of 
application migration. On the other hand, many works have shown 
that extracting a  large fraction of the performance capabilities 
of the Knights Corner (KNC),  the first generation Xeon Phi 
co-processors, still requires significant efforts and fine restructuring of the code.

Following recent improvements in directive-based programming environments, the
present work wants to explore ways of writing  common codes that i) have
optimization features that can be exploited  by traditional CPUs and both accelerator architectures,
and ii) are written using a common  directive-based (e.g. OpenMP, OpenACC)
programming environment.

%%%%%%%%%%%%%%%%%%%%%%%%%%%%%%%%%%%%%%%%%%%%%%%%%%%%%%%%%%%%%%%%%%%%%%

\section{Lattice Boltzmann methods}\label{sec:lbm}

In this section, we sketchily introduce the computational method that we adopt,
based on an advanced Lattice Boltzmann (LB) scheme. LB methods (see
~\cite{sauro} for an introduction) are discrete in position
and momentum spaces; they are based on the synthetic dynamics of 
{\em populations} sitting at the sites of a discrete lattice. At each time step,
populations hops from lattice-site to lattice-site and then they
{\em collide}, mixing and changing their values
accordingly.

Over the years, several LB models have been developed,
describing flows in 2 or 3 dimensions, and using sets of populations of
different size (a model in $x$ dimensions based on $y$ populations is
labeled as $DxQy$). Populations ($f_l(x,t)$), each having a given lattice 
velocity $\bm c_{l}$, are defined at the sites of a discrete and regular grid;
they evolve in (discrete) time according to the Bhatnagar-Gross-Krook (BGK) equation:
\begin{align}\label{eq:lbequation}
& f_{l}({\bm x}+ {\bm c}_{l} \dt,t+\dt) = \nonumber \\ 
& f_{l} ({\bm x},t) -\frac{\dt}{\tau}\left(f_{l}({\bm x},t) - f_l^{(eq)}\right)
\end{align} 
The macroscopic physics variables, density $\rho$, velocity $\bm u$  and
temperature $T$ are defined in terms of the
$f_l(x,t)$ and $\bm c_{l}$ variables:
\begin{align}\label{eq:macro}
\rho = \sum_l f_l && \rho {\bm u} = \sum_l {\bm c}_l f_l && D \rho \T = \sum_l \left|{\bm c}_l - {\bm u}\right|^2 f_l ; \nonumber
\end{align}
the equilibrium distributions (${f}_{l}^{(eq)}$) are themselves functions of
these macroscopic quantities (\cite{sauro}). 
With an appropriate choice of the
set of lattice velocities $\bm c_{l}$ and of the  equilibrium distributions
${f}_{l}^{(eq)}$, one shows that, performing an expansion in $\Delta t$ and
re-normalizing the values of the physical velocity and temperature fields,
the evolution
of the macroscopic variables obeys the thermo-hydrodynamical equations of
motion and the continuity equation:
\begin{align}
& \partial_t \rho + \rho \pii u_i = 0, \nonumber \\
& \rho D_t u_i = - \pii p - \rho g\delta_{i,2} + \nu \partial_{jj} u_i, \\
& \rho c_v D_t T + p \pii u_i = k \partial_{ii} T; \nonumber
\end{align}
$D_t = \partial_t + u_j \pj$ is the material derivative and 
we neglect viscous heating;
$c_v$ is the specific heat at constant volume for an ideal gas, 
$p=\rho T$, and $\nu$ and $k$ are the transport coefficients; $g$ is the
acceleration of gravity, acting in the vertical direction. Summation of
repeated indexes is implied.

In our case we study a 2-dimensional system ($D=2$ in the following), 
and the set
of populations has $37$ elements (hence the D2Q37 acronym) corresponding to
(pseudo-)particles moving up to three lattice points away, as shown in 
\autoref{fig:streamscheme} (\cite{JFM,POF}).
The main advantage of this recently developed LB method is
that it automatically enforces the equation of state of a perfect gas 
($p = \rho T$).
Our optimization efforts have made it possible to perform large scale simulations 
of convective turbulence in several physics conditions (see e.g.,~\cite{noi1,noi2}); 

An LB code starts with an initial assignment of the populations, corresponding to 
a given initial condition at $t = 0$ on some spatial domain, and
iterates \autoref{eq:lbequation} for each population and lattice site and for
as many time steps as needed; boundary conditions are
enforced at the edges of the domain after each time step
by appropriately modifying population values at and close to the boundary.

From the computational point of view, the LB approach offers a
huge degree of easily identified available parallelism.
Defining ${\bm y} = {\bm x}+ {\bm c}_{l} \dt$ and rewriting the main
evolution equation as:
\begin{align}\label{eq:master2}
& f_{l}({\bm y}, t+\dt) = \nonumber \\
& f_{l}({\bm y} - {\bm c}_{l} \dt,t) - \frac{\dt}{\tau}\left(f_{l}({\bm y} - {\bm c}_{l} \dt,t) - f_l^{(eq)}\right)
\end{align}
one easily identifies the overall structure of the computation that evolves the
system by one time step $\dt$: for each point ${\bm y}$ in the discrete grid the code: 
i) gathers from neighboring sites the values of the fields $f_l$ corresponding to 
populations  drifting towards ${\bm y}$ with velocity
${\bm c}_l$ and then, ii) performs all mathematical steps needed to compute the 
quantities in the r.h.s. of \autoref{eq:master2}. 
One quickly sees that there is no correlation between different
lattice points, so both steps can proceed in parallel on all grid points
according to any convenient schedule, with the only constraint that step 1
precedes step 2. 

As already remarked, our D2Q37 model correctly and consistently describes the
thermo-hydrodynamical equations of motion and the equation of state of a perfect
gas; the price to pay is that, from a computational point of view, its
implementation is more complex than for simpler LB models.  This translates
in demanding requirements for memory bandwidth and
floating-point throughput. Indeed, step 1 implies accessing 37 neighbor cells to
gather all populations, while step 2 implies $\approx 7000$ double-precision
floating point operations per lattice point, some of which can be optimized away
e.g. by the compiler.

%%%%%%%%%%%%%%%%%%%%%%%%%%%%%%%%%%%%%%%%%%%%%%%%%%%%%%%%%%%%%%%%%%%%%%

\section{Data-layout optimization for LB kernels}\label{sec:implementation}

Our goal is to design a performance-portable code capable of running
efficiently on recent Intel multi-core CPUs as well as on Xeon-Phi and GPU
accelerators. 

Our intendedly common application is written in plain {\tt C} and  annotated
with compiler directives for parallelization. For Intel architectures, we have
used OpenMP and proprietary Intel directives, using the offload pragmas to run 
kernels on the Xeon-Phi. 
For NVIDIA and AMD GPUs we have annotated the code with OpenACC directives, 
implemented by the PGI compiler, which supports both architectures. 
Mapping OpenACC directives on OpenMP directives is almost  straightforward, so
code divergence is limited at this point in time; it is expected that OpenMP
implementations supporting both classes of accelerators will become 
available in the near future, so we hope to be able soon to merge the two 
versions into one truly common code.

%A critical point to reach this goal is the design 
%of appropriate data layouts for the critical kernels of our D2Q37 LB code --
%{\tt propagate}  and {\tt collide} -- able to efficiently exploit both
%architectures.
As remarked in \autoref{sec:introduction}, data layout 
has a critical role in extracting performance from accelerators.
% 
%We start our discussion and analysis on Intel architectures, multi-core 
%CPUs and many-core accelerators, and then we extend 
%our result to NVIDIA and AMD GPUs.

%%%%%%%%%%%%%%%%%%%%%%%%%%%%%%%%%%%%%%%%%%%%%%%%%%%%%%%%%%%%%%%%%%%%%%
\begin{figure}[t]
\centering
\begin{minipage}{0.48\textwidth}
\begin{lstlisting}[basicstyle=\scriptsize,language=C]
#define N (LX*LY)
typedef struct {
  data_t  p1; // population  1 
  data_t  p2; // population  2 
  ...
  data_t p37; // population 37 
} pop_t;

pop_t lattice[N];
\end{lstlisting}
\end{minipage}
%
%\hspace*{2.5mm}
%
\begin{minipage}{0.48\textwidth}
\begin{lstlisting}[basicstyle=\scriptsize,language=C]
#define N (LX*LY)
typedef struct {
  data_t  p1[N]; // population  1 
  data_t  p2[N]; // population  2 
  ...
  data_t p37[N]; // population 37 
} pop_t;

pop_t lattice;
\end{lstlisting}
\end{minipage}
\vskip 2.5mm
\includegraphics[width=0.49\textwidth]{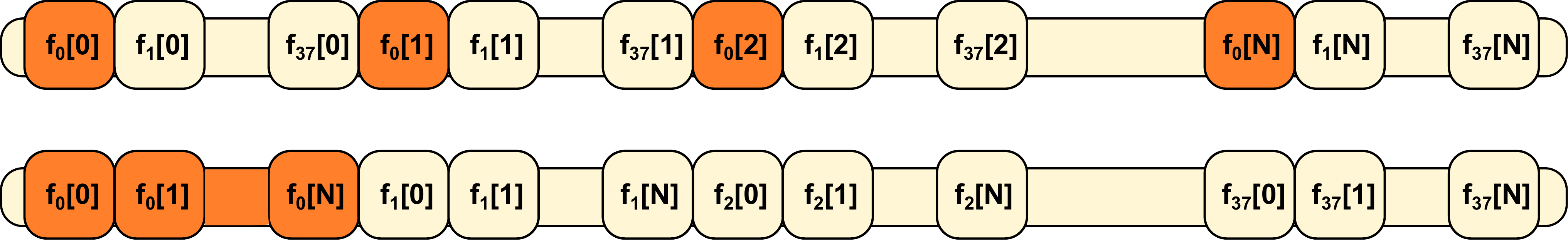}
\caption{
Top and Middle: {\tt C} definitions of lattice populations using the AoS 
and SoA data layouts.
%In the AoS layout, populations belonging to the same lattice site are 
%stored at contiguous memory locations while populations of common index $i$ 
%are stored at non unit-stride addresses. Conversely, the SoA layout
%stores contiguously all populations of index $i$ from different lattice sites.
Bottom: Graphic representation of the AoS and SoA layout.
}\label{code:aos-vs-soa}
\end{figure}
%%%%%%%%%%%%%%%%%%%%%%%%%%%%%%%%%%%%%%%%%%%%%%%%%%%%%%%%%%%%%%%%%%%%%%

Data layouts for LB methods, as for many other stencil applications, 
have been traditionally based either on {\em array of structures} (AoS) or on 
{\em structure of arrays} (SoA) schemes. 
\autoref{code:aos-vs-soa} shows how population data are stored in the 
two cases;
in the AoS layout, population data for each lattice site are stored one after 
the other at neighboring memory locations.
This scheme exhibits locality of populations at a given lattice site, while 
populations of common index $i$ at different lattice sites are stored in 
memory at non unit-stride addresses.
Conversely, in the SoA scheme for a given index $i$, populations of all lattice sites 
are stored contiguously, while the various populations 
of each lattice site are stored far from each other at non unit-stride addresses. 

In our implementation we have arbitrarily chosen to store the lattice 
in column-major order ($Y$ spatial direction); we keep in memory two copies 
that are alternatively read and written by each kernel routine.
Although demanding in terms of memory requirements, this choice has 
the key advantage that it allows in principle to process all lattice sites in parallel.

%%%%%%%%%%%%%%%%%%%%%%%%%%%%%%%%%%%%%%%%%%%%%%%%%%%%%%%%%%%%%%%%%%%%%%%%

\subsection{Data-layout optimization for {\tt propagate}}\label{subsec:propagate}

In our LB code the {\tt propagate} kernel is applied to each lattice site, 
moving populations according to the patterns of~\autoref{fig:streamscheme}.
For each site, {\tt propagate} reads and writes 
populations from lattice cells at distance up to 3 in the physical lattice 
and for this reason locality of memory accesses plays an important role 
for performance. 

 This kernel can be implemented using either a {\em push} or a {\em pull} 
scheme,~\cite{Wittmann}. The former moves all populations of a lattice site 
towards appropriate neighboring sites, while the latter gathers to one site 
populations stored at neighbor sites. 
Relative advantages and disadvantages of these schemes are not obvious 
and depend to some extent on the hardware features of the target processor. 
While the {\em push} scheme performs an aligned-read followed by a misaligned-write, 
the opposite happens if the {\em pull} scheme is used, and it is well known that 
reading/writing from/to (non-)aligned memory addresses may have a large impact 
on the sustained memory bandwidth of modern processors,~\cite{sbac-pad13}. 
However, on cache-based Intel architectures (both standard CPUs and Xeon-Phi), 
aligned data can be stored directly to memory using {\em non-temporal} 
write instructions. 
If data to be stored is not resident at any cache level, standard semantic of ordinary 
memory-writes require a prior read for ownership (RFO); the {\em non-temporal} 
version of store avoids the RFO read, improving effective memory 
bandwidth and saving time. 
This feature can be used in the {\em pull} scheme reducing the overall 
memory traffic by a factor $1/3$, so we adopt it.

%%%%%%%%%%%%%%%%%%%%%%%%%%%%%%%%%%%%%%%%%%%%%%%%%%%%%%%%%%%%%%%%%%%%%%
%
\begin{figure*}[t]
\centering
\begin{minipage}[t]{0.48\textwidth}
\begin{lstlisting}[basicstyle=\scriptsize,language=C]
typedef data_t double;
typedef struct { data_t s[LX*LY]; } pop_soa_t;

pop_soa_t prv[NPOP], nxt[NPOP];
 
 
#pragma omp parallel for
for ( ix = STARTX; ix < ENDX; ix++ ) {
  #pragma vector nontemporal
  for ( iy = STARTY; iy < ENDY; iy++ ) {
    idx = IDX(ix, iy);
    for ( ip = 0; ip < NPOP; ip++ ) {
      nxt[ip].s[idx] = prv[ip].s[ idx + OFF[ip] ];
    }
  }
}  
\end{lstlisting}
\end{minipage}
\hspace*{1mm}
\begin{minipage}[t]{0.48\textwidth}
\begin{lstlisting}[basicstyle=\scriptsize,language=C]
typedef data_t double;
typedef struct { data_t s[LX*LY]; } pop_soa_t;

pop_soa_t prv[NPOP], nxt[NPOP];
 
#pragma acc kernels present(prv, nxt)
#pragma acc loop gang independent
for ( ix = STARTX; ix < ENDX; ix++ ) {
  #pragma acc loop vector independent
  for ( iy = STARTY; iy < ENDY; iy++ ) {
    idx = IDX(ix, iy);
    for ( ip = 0; ip < NPOP; ip++ ) {
      nxt[ip].s[idx] = prv[ip].s[ idx + OFF[ip] ];
    }
  }
}  
\end{lstlisting}
\end{minipage}
\caption{ 
Codes of the {\tt propagate} kernel for Intel architectures (left) and GPUs (right). 
This kernel moves populations as shown in \autoref{fig:streamscheme}. 
{\tt OFF} is a vector containing the memory address offsets associated to each 
population hop. 
The {\tt prv} and {\tt nxt} arrays use the SoA layout. 
%Structures of type {\tt pop\_soa\_t} store each a population array.
}\label{code:propagate-soa}
\end{figure*}
%
%%%%%%%%%%%%%%%%%%%%%%%%%%%%%%%%%%%%%%%%%%%%%%%%%%%%%%%%%%%%%%%%%%%%%%

The {\tt propagate} kernel can in principle be vectorized
applying each move shown in~\autoref{fig:streamscheme} to several 
lattice sites in parallel, e.g. moving populations with the same index $i$ 
and belonging to two or more sites.  
% FINO QUI!!!!
The number of sites processed in parallel depends on the size of the 
vector instructions of the target processor; vector size 
is $4$ double words for the Xeon-CPU, $8$ for the Xeon-Phi, $32$ 
or multiples thereof for GPUs.
However, using the AoS scheme, populations of different sites are stored at 
non-contiguous memory addresses preventing vectorization.
Conversely, when using the SoA layout, access to several populations of index $i$ 
has unit stride, allowing to move in parallel populations of 
index $i$ for several sites and allowing vectorization.  
This discussion suggests that the SoA layout should perform better than the AoS one.
\autoref{code:propagate-soa} (left) shows the C-code written for Intel 
processors, adopting the SoA layout.
The code sweeps all lattice sites with two loops in the $X$ and $Y$ spatial directions. 
The inner loop is on $Y$, as elements are stored in column-major order.  
We have annotated this loop with the {\tt \#pragma omp simd} OpenMP pragma 
to introduce SIMD vector instructions. 
Since values of populations written into {\tt nxt} array are not re-used within this kernel, 
we also enable {\em non-temporal} stores; since this feature is not yet part of the
OpenMP standard we have used the specific Intel directive {\tt \#pragma vector nontemporal}. 
The equivalent code for GPUs, replacing Intel and OpenMP directives with 
corresponding OpenACC ones, is shown in \autoref{code:propagate-soa} (right). 

%%%%%%%%%%%%%%%%%%%%%%%%%%%%%%%%%%%%%%%%%%%%%%%%%%%%%%%%%%%%%%%%%%%%%%
\begin{table}
\centering 
\begin{adjustbox}{width=0.49\textwidth}
\begin{tabular}{l r r r r} 
\toprule 
Data Structure & Haswell & Xeon Phi & Tesla K80 & AMD Hawaii \\ 
\midrule
\multicolumn{5}{c}{propagate} \\
\midrule
AoS        & 408  & 194   &   326  & 649 \\
SoA        & 847  & 224   &    36  &  57 \\
CSoA       & 247  &  78   &    32  &  45 \\
CAoSoA     & 286  &  89   &    33 &  50  \\ 
\midrule
\multicolumn{5}{c}{collide} \\
\midrule
AoS        & 1232 &  631 &  767    & 2270 \\
SoA        & 1612 & 1777 &  171    & 1018 \\
CSoA       &  955 &  445 &  165    &  452 \\
CAoSoA     &  812 &  325 &  166    &  402 \\
\bottomrule
\end{tabular} 
\end{adjustbox}
\caption{
Execution time (milliseconds per iteration) of the {\tt propagate} and {\tt collide} 
kernels on several architectures using different data-layouts. The size of the lattice 
is $2160 \times 8192$ points.
%% (*) Results on AMD got using PGI 16.5 and Driver AMD May 2016.
}\label{tab:propagate-kernel}\label{tab:collide-kernel}
\end{table}
%%%%%%%%%%%%%%%%%%%%%%%%%%%%%%%%%%%%%%%%%%%%%%%%%%%%%%%%%%%%%%%%%%%%%%
 
The first two rows of \autoref{tab:propagate-kernel} compare the performance 
obtained with the two layouts on all the processors that we consider. As expected 
the SoA layout is much more efficient on GPUs, but this is not true for both 
Intel processors; inspection of the assembly codes and compiler logs shows that 
read operations are not vectorized on Intel processors due to {\em unaligned} 
load addresses. Lacking vectorization, the AoS layout exhibits a better performance 
as it has a better cache hit rate.

The reason why compilers fail to vectorize the code is that load addresses 
are computed as the sum of the destination-site address -- which is memory-aligned -- 
and an offset, so they point to the neighbor sites from which populations are read. 
This does not guarantee that the resulting address is properly aligned 
to the vector size, i.e. 32 Bytes for CPUs and 64 Bytes for the Xeon-Phi.
Store addresses on the other hand are always aligned if the lattice 
base address is properly aligned and $Y$ is a multiple of 32 or 64.

A simple modification to the data layout solves this problem. 
Starting from the lattice stored in the SoA scheme, we cluster {\tt VL} consecutive 
elements of each population array, with {\tt VL} a multiple of the hardware 
vector size supported by the processor (e.g. 4 for the Haswell CPU, 
8 for the Xeon-Phi and 32 for GPUs). 
We call this scheme {\em Cluster Structure of Array} (CSoA);
in \autoref{code:propagate-csoa} we show the corresponding 
{\tt C} type definitions, {\tt vdata\_t} and {\tt vpop\_soa\_t}.
{\tt vdata\_t} holds {\tt VL} data words corresponding to the 
same population of index $i$ at {\tt VL} different sites that can be 
processed in SIMD fashion. 
{\tt vpop\_soa\_t} is the type definition for the full lattice data.
Using this scheme, move operations generated by {\tt propagate} apply 
to clusters of populations and not to individual population elements. Since 
clusters have the same size as hardware vectors, all read operations are now 
properly aligned. 
As in the case of the SoA data layout, write operations always have 
aligned accesses and non temporal stores can be used.
%
%%%%%%%%%%%%%%%%%%%%%%%%%%%%%%%%%%%%%%%%%%%%%%%%%%%%%%%%%%%%%%%%%%%%%%
\begin{figure}[t]
\centering
\begin{lstlisting}[basicstyle=\scriptsize,language=C]
#define LYOVL (LY / VL)
typedef data_t double;
typedef struct { data_t c[VL]; } vdata_t;

typedef struct { vdata_t s[LX*LYOVL]; } vpop_soa_t;

vpop_soa_t prv[NPOP], nxt[NPOP];

for ( ix = startX; ix < endX; ix++ ) {
  idx = ix*LYOVL;
  for( ip = 0; ip < NPOP; ip++) {
    for ( iy = startY; iy < endY; iy++ ) {
      #pragma vector aligned nontemporal
      for(k = 0; k < VL; k++) {
        nxt[ip].s[idx + iy].c[k] = 
	      prv[ip].s[idx + iy + OFF[ip]].c[k];
      }
    }
  }
}
\end{lstlisting}
\caption{
Source code of the {\tt propagate} kernel for Intel architectures using 
the CSoA data layout.
{\tt OFF} is a vector containing the memory-address offsets associated 
to each population hop. 
{\tt VL} is the size of a cluster (see text for details). 
To properly vectorize the inner loop with SIMD instructions, 
value of {\tt VL} should match the width of vector-registers supported 
by the target architecture. 
}\label{code:propagate-csoa}
\end{figure}
%%%%%%%%%%%%%%%%%%%%%%%%%%%%%%%%%%%%%%%%%%%%%%%%%%%%%%%%%%%%%%%%%%%%%%
%
\autoref{code:propagate-csoa} shows the corresponding code; 
in this case we have also rearranged the order of the loops in a way which 
reduces the pressure on the TLB cache.
As before, code for GPUs can be obtained replacing directives with 
{\em OpenAcc} ones.

\autoref{tab:propagate-kernel} (see first three rows of the {\tt propagate} section) 
quantifies the impact of the data layout 
on performance, showing benchmark results using the three
different data layouts, AoS, SoA and CSoA. 
The advantages of using the CSoA data layout are large for Intel
architectures, while improvements are marginal  for GPUs, as they are  less
sensitive to misaligned memory reads~\cite{sbac-pad13}.  
The relevant result is however that using the CSoA format 
we have one common data layout that maximizes performance for 
{\tt propagate} on all processors.

%%%%%%%%%%%%%%%%%%%%%%%%%%%%%%%%%%%%%%%%%%%%%%%%%%%%%%%%%%%%%%%%%%%%%%

%% %%%%%%%%%%%%%%%%%%%%%%%%%%%%%%%%%%%%%%%%%%%%%%%%%%%%%%%%%%%%%%%%%%%%%%
%% \begin{table}
%% \centering
%% \begin{adjustbox}{width=0.49\textwidth}
%% \begin{tabular}{l cc}
%% \toprule
%% Metric  	            & VTune-output & Threshold \\
%% \midrule
%% Xeon-Phi L1 TLB Miss Ratio  & 2.66\%       & 1.0\%     \\
%% Xeon-Phi L2 TLB Miss Ratio  & 2.00\%       & 0.1\%     \\
%% \midrule
%% Xeon-CPU LLC Miss Count     & 787,647,256  & n/a       \\
%% \bottomrule       
%% \end{tabular}
%% \end{adjustbox}
%% \caption{ 
%% Profiling results provided by Intel VTune for the {\tt collide} 
%% kernel using the CSoA scheme on a lattice of $2160\times8192$ 
%% points for the Xeon-CPU and Xeon-Phi processors.
%% }\label{tab:csoa-profile}
%% \end{table}
%% %%%%%%%%%%%%%%%%%%%%%%%%%%%%%%%%%%%%%%%%%%%%%%%%%%%%%%%%%%%%%%%%%%%%%%

\subsection{Data-layout optimization for {\tt collide}}\label{subsec:collide}

The {\tt collide} kernel is computed after the {\tt propagate} step, reading 
at each lattice site populations gathered by the {\tt propagate} phase. 
It updates their values applying the {\em collisional} operator and performing 
all mathematical operations associated with \autoref{eq:lbequation}. 
For each lattice site, this floating point intensive kernel uses only 
population data associated to the site on which it operates; lattice sites 
are processed independently from each other making processing of the lattice 
fully parallelizable.
In this case, in contrast with the {\tt propagate} kernel, locality 
of populations plays an important role for performances.
Vectorization of this step is implemented, as for {\tt propagate}, 
trying to process different sites in parallel. 

Following results obtained in \autoref{subsec:propagate} we consider 
first the CSoA data layout which -- as seen before -- 
gives very good performance results with the {\tt propagate} kernel.
% 
% Using the CSoA scheme, the compiler is able to correctly vectorize 
% the loops within the collide function, but the code exhibits poor 
% performance both on Xeon-CPU and Xeon-Phi. 
%
% \todo{A: Non e' vero che abbiamo scarse performance, anzi, CSoA da significativi
% benefici sia su host che device (vedi tabella). Possiamo dire che guardando le
% metriche del TLB capiamo che c'e' ancora spazio per ulteriori miglioramenti.}
%
The log files of the compiler show that the CSoA scheme allows  
to vectorize the code, but profiling the execution of the code on Xeon CPUs 
and Xeon Phi accelerators, we have observed a large number of TLB and 
LLC misses, suggesting that further improvements could be put in place. 
\autoref{tab:csoa-profile} shows the results (see CSoA column) provided by 
Intel VTune profiler for both Xeon-CPU and Xeon-Phi; for the latter the 
miss ratios are much bigger than threshold values provided by the profiler itself.    
These penalties arise because in the CSoA scheme different 
populations associated to the same lattice site are stored at 
memory addresses far from each other, so several 
non-unit stride memory accesses are necessary to gather all relevant 
data words.

%%%%%%%%%%%%%%%%%%%%%%%%%%%%%%%%%%%%%%%%%%%%%%%%%%%%%%%%%%%%%%%%%%%%%%
\begin{table}
\centering
\begin{adjustbox}{width=0.49\textwidth}
\begin{tabular}{l r r r }
\toprule
                  Metric &  CSoA & CAoSoA & Threshold  \\
\midrule
\multicolumn{4}{c}{Xeon-Phi} \\
\midrule
L1 TLB Miss Ratio        &      2.66\% &       0.06\% &      1.0\% \\
L2 TLB Miss Ratio        &      2.00\% &       0.00\% &      0.1\% \\
\midrule
\multicolumn{4}{c}{Xeon-CPU} \\
\midrule
LLC Miss Count           & 787,647,256 &  177,010,620 &       n/a  \\
Average Latency (cycles) &          13 &            9 &       n/a  \\
\bottomrule
\end{tabular}
\end{adjustbox}
\caption{
Profiling results provided by the Intel VTune profiler for the {\tt collide}
kernel on a lattice of $2160\times8192$ points comparing the CSoA
and the CAoSoA scheme on the Xeon-CPU and Xeon-Phi processors.
}\label{tab:csoa-profile}\label{tab:caosoa-profile}
\end{table}
%%%%%%%%%%%%%%%%%%%%%%%%%%%%%%%%%%%%%%%%%%%%%%%%%%%%%%%%%%%%%%%%%%%%%%

\begin{figure}
\centering
\includegraphics[width=0.49\textwidth]{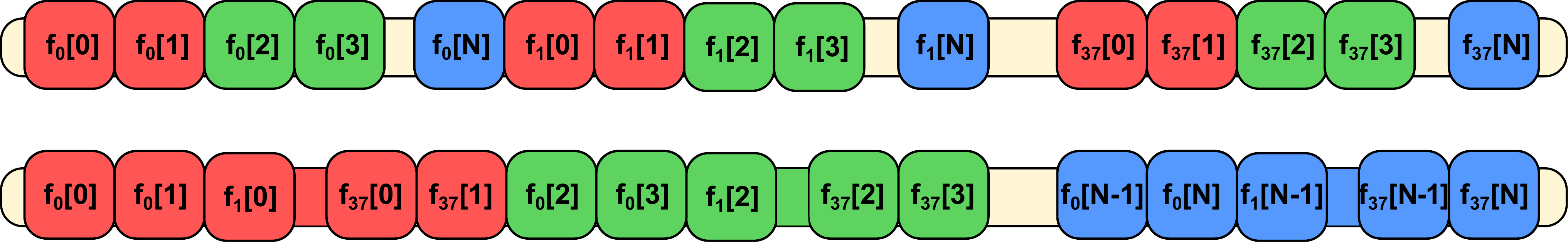}
\vspace*{1mm}
\begin{lstlisting}[basicstyle=\scriptsize,language=C]
#define LYOVL (LY / VL)
// cluster definition
typedef struct { double c[VL]; } vdata_t;  

// CAoSoA type definition
typedef struct { vdata_t p[NPOP]; } caosoa_t; 

vpop_soa_t prv[LX*LYOVL], nxt[LX*LYOVL];

// snippet of collide code to compute density rho
vdata_t rho;
for (ip = 0; ip < NPOP; ip++)
  #pragma vector aligned
  for (k = 0; k < VL; k++)
    rho.c[k] = rho.c[k] + prv[idx].p[ip].c[k];
\end{lstlisting}
\caption{ Top: data arrangement for the CAoSoA layout; 
for illustration purposes, we take VL=2.
Bottom: sample code for {\tt collide} using this layout.
}\label{fig:caosoa-scheme}
\end{figure}

To overcome these problems, we further modify the CSoA scheme, and define 
a new data layout which takes into account locality requirements of 
both {\tt propagate} and {\tt collide} kernels.
In this scheme, for each population array, we divide each $Y$-column 
in {\tt VL} partitions each of size {\tt LY/VL}; all elements sitting at the
$i$th position of each partition are then packet together into an array of 
{\tt VL} elements called {\em cluster}.
We call this layout a {\em Clustered Array of Structure of Array} (CAoSoA) 
and \autoref{fig:caosoa-scheme} shows how data are arranged in memory.
This data layout still allows vectorization of inner structures (clusters) 
of size VL, and at the same time improves locality -- w.r.t the CSoA -- 
of populations, as it keeps all population data 
needed to process each lattice site at close and aligned addresses.
\autoref{fig:caosoa-scheme} shows  the definition of the {\tt vdata\_t} 
data type, corresponding to a {\em cluster}, and a representative 
small sections of the {\tt collide} code for Intel processors. 
%
%%% The corresponding code, annotated with OpenACC directives that we have 
%%% used for GPUs is shown in \autoref{code:collide-openacc}.
% 
Cluster variables are processed iterating on all elements of the cluster 
through a loop over VL; {\tt pragma vector aligned} instructs the 
compiler to fully vectorize the loop since all iterations are independent 
and memory accesses are aligned.
This data-layout combines the benefits of the CSoA scheme, allowing 
aligned memory  accesses and vectorization (relevant for the {\em propagate} kernel)
and at the same providing population locality ( together relevant 
for the {\em collide} kernel). 

%%%%%%%%%%%%%%%%%%%%%%%%%%%%%%%%%%%%%%%%%%%%%%%%%%%%%%%%%%%%%%%%%%%%%%
%
%
%%%%%%%%%%%%%%%%%%%%%%%%%%%%%%%%%%%%%%%%%%%%%%%%%%%%%%%%%%%%%%%%%%%%%%

\autoref{tab:csoa-profile} shows the impact of the CAoSoA  
data-layout on memory misses: on Xeon-Phi the TLB misses have 
been reduced well below the threshold values, and on Xeon-CPU 
have been reduced by a factor 4.5X w.r.t. the CSoA scheme.
\autoref{tab:collide-kernel} shows the execution time of the 
{\tt collide} kernel run using all data data-layouts 
defined so far.
As wee see, the CAoSoA improves performances over the CSoA 
on Intel and AMD processors while for NVIDIA GPU the two layouts give marginal 
differences in performance.
These gains in the performance of {\tt collide} come at a limited 
cost ($12-16\%$) for {\em propagate} on all architectures except for the 
K80, so CAoSoA maximizes the combined performances of the two kernels. 

%%%%%%%%%%%%%%%%%%%%%%%%%%%%%%%%%%%%%%%%%%%%%%%%%%%%%%%%%%%%%%%%%%%%%%
\section{Heterogeneous implementation}\label{sec:heterogeneous}

In this section we describe the implementation of our code designed to 
involve and exploit compute capabilities of both host and accelerators. 
We only consider the CAoSoA layout, as it grants the best 
overall performances on all processors and accelerators that we have studied. 

%%%%%%%%%%%%%%%%%%%%%%%%%%%%%%%%%%%%%%%%%%%%%%%%%%%%%%%%%%%%%%%%%%%%%%
\begin{figure}
\centering
\includegraphics[width=.49\textwidth]{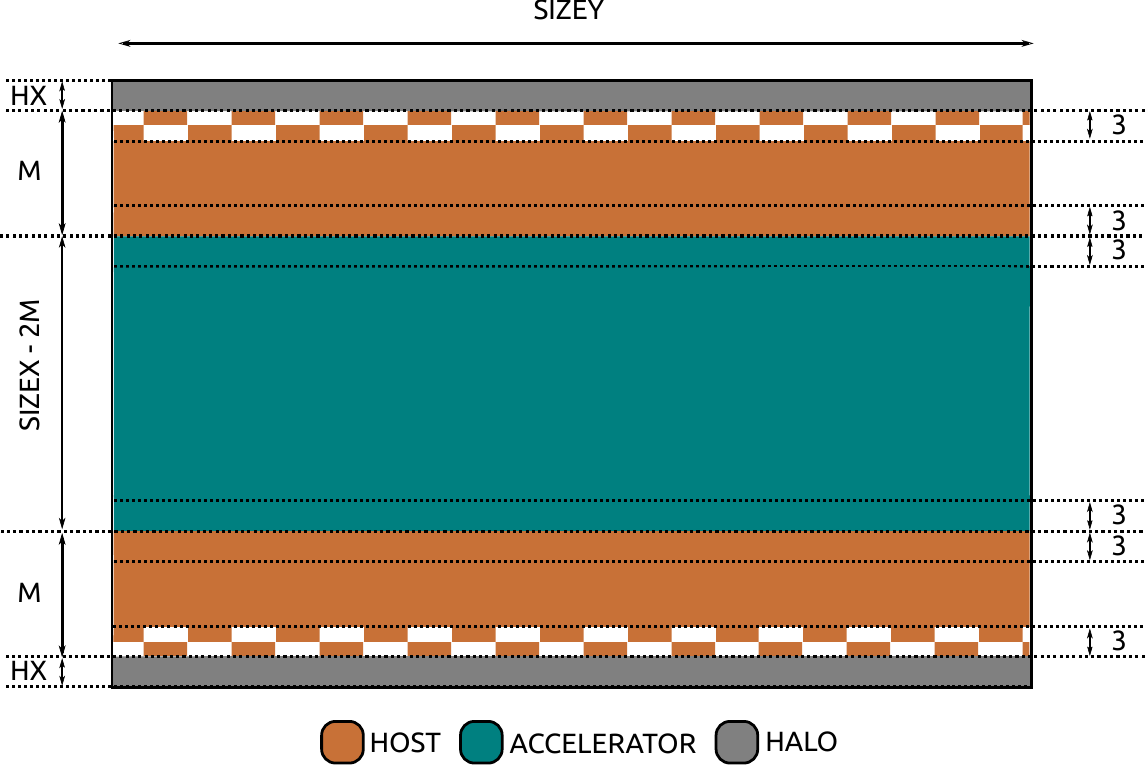}
\caption{
Logic partitioning of the lattice domain among host and accelerator.
The central (dark-green) region is allocated on the accelerator, 
side (orange and gray) regions on the host. Checkerboard 
textures flag lattice regions involved in MPI communications with 
neighbor nodes.
}\label{fig:lattice-het}
\end{figure}
%%%%%%%%%%%%%%%%%%%%%%%%%%%%%%%%%%%%%%%%%%%%%%%%%%%%%%%%%%%%%%%%%%%%%%

Our implementation uses MPI libraries and each MPI-process manages 
one accelerator. The MPI-process runs on the host CPU; part 
of the lattice domain is processed on the host itself, and part is 
offloaded and processed by the accelerator. 
Using one MPI-process per accelerator makes it easy to extend the 
implementation to a cluster of accelerators installed 
either on the same or on different hosts. 

The lattice is partitioned among the MPI-processes, along one direction, 
in our case the $X$-direction, and each slice is assigned to a different 
MPI-process.
Within each MPI-process each partition is further divided between host and 
accelerator. We define three regions, namely 
left border, bulk, and right border, as shown in \autoref{fig:lattice-het}. 
The right and left borders include $M$ columns and are allocated on the host 
memory while the remaining {\tt SIZEX - $2M$} columns stay on the accelerator memory. 
As {\tt propagate} stencils require to access neighbor sites 
at distance up to 3, see \autoref{fig:streamscheme}, each region 
is surrounded by a halo of three columns and rows. 
Each halo stores a copy of lattice sites of the adjacent region either 
allocated on the host, on the accelerator, or on the neighbor 
MPI-process. 
This arrangement allows to uniformly apply the {\tt propagate} kernel 
to all sites avoiding divergences in the computation.
An added advantage of this layout is that data involved in MPI 
communications is resident on the host and not on the accelerators, 
slightly increasing data locality.
%
%%%%%%%%%%%%%%%%%%%%%%%%%%%%%%%%%%%%%%%%%%%%%%%%%%%%%%%%%%%%%%%%%%%%%%
%
\begin{figure}
\includegraphics[width=.49\textwidth]{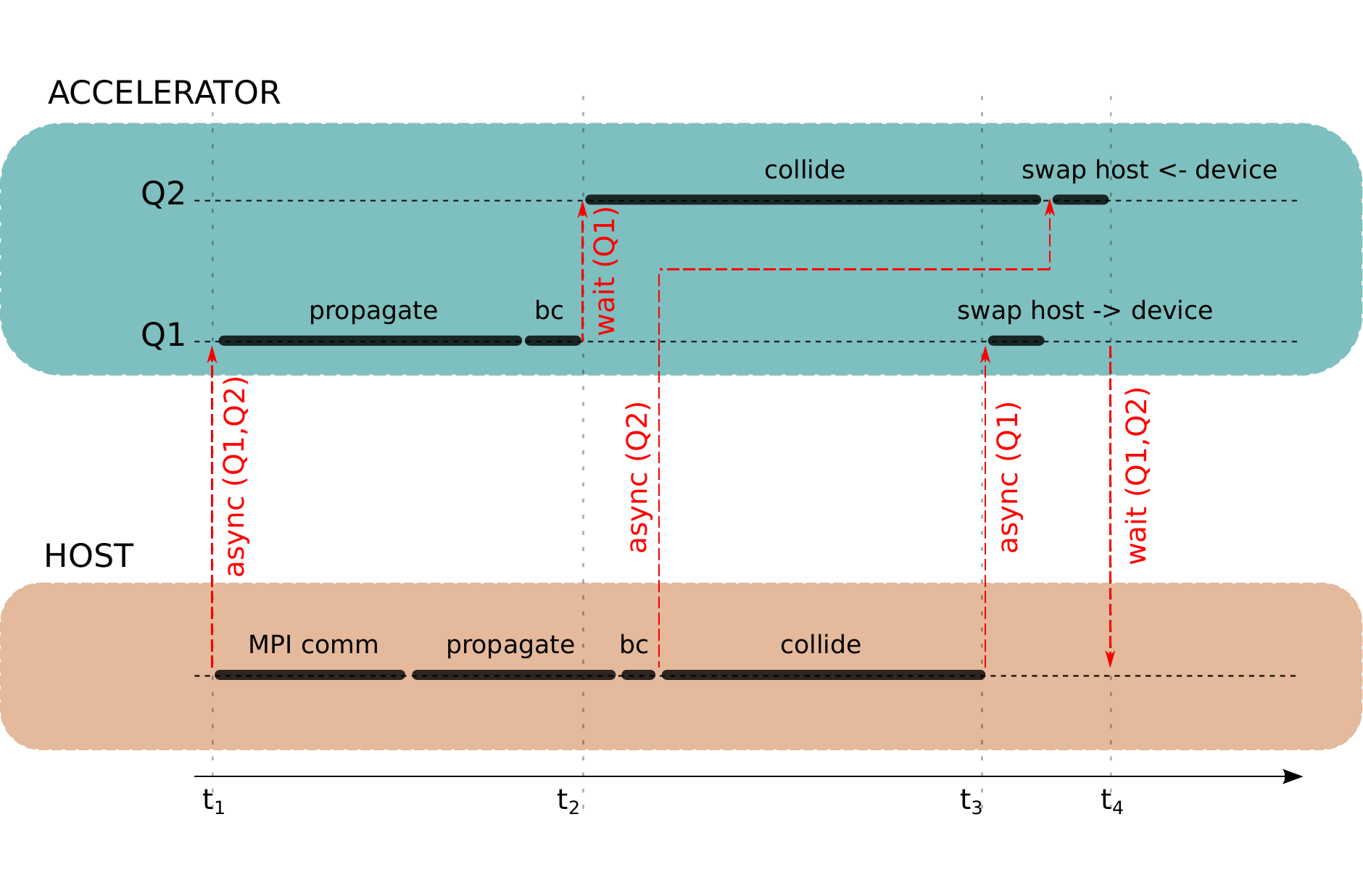}
\caption{
Control flow executed by each MPI-process. 
The schedule executed on the accelerator is on the upper band, while the one
executed by the host is on the lower band. 
Execution on the accelerator runs on two concurrent 
queues. Synchronization-points are marked 
with red lines.
}\label{fig:control-flow}
\end{figure}
%
%%%%%%%%%%%%%%%%%%%%%%%%%%%%%%%%%%%%%%%%%%%%%%%%%%%%%%%%%%%%%%%%%%%%%%

Each MPI-process performs a loop over time steps, and at each iteration 
it launches in sequence on the accelerator the {\tt propagate}, 
{\tt bc} and {\tt collide} kernels, processing the bulk region. 
In order to allow the CPU to operate in overlapped mode, kernels are launched 
asynchronously on the same logical execution queue to ensure in-order execution.
After launching the kernels on the accelerator, the host first updates halos with 
adjacent MPI-processes and then starts to process its left and right borders 
applying the same sequence of kernels. 
After the processing of borders completes, the host updates the halo 
regions shared with the accelerator; this step moves data between 
host and accelerator. 
The control-flow of the code executed by the MPI-process is shown 
in \autoref{fig:control-flow}, where the {\tt bc} kernel applies the physical 
boundary conditions at the three uppermost and lowermost rows of the lattice.  

%
%Boundary and bulk regions are processed in parallel respectively by host-CPU 
%and accelerator, but at each iteration an exchange of data between them 
%is required to update halo-regions.
%

The code for KNC is implemented using the offload features available in the Intel 
compiler and runtime framework. 
In this case we have a unique code and the compiler produces the executable 
codes for both Xeon-Phi and CPU.
For GPUs the situation is somewhat different. In fact, we have to use 
two different compilers, one for GPUs and one for CPUs. We have written the 
code for GPUs, both NVIDIA and AMD, using {\em OpenACC} directives~\cite{ccpe16} 
and compiled using the PGI~16.5 compiler which supports both architectures. 
The kernels running on CPUs are written using standard {\tt C} and compiled 
using Intel compiler ICC v16.0. 
Then we have linked the two codes in one single executable.

%%%%%%%%%%%%%%%%%%%%%%%%%%%%%%%%%%%%%%%%%%%%%%%%%%%%%%%%%%%%%%%%%%%%%%
\section{Parameter optimization}\label{sec:optimization}

In this section we describe two important optimization steps for our 
heterogeneous code. 
The first is about the optimal partitioning of the computation between host 
and accelerator, and the latter is about the optimal cluster size  
for the CAoSoA data layout.

%%%%%%%%%%%%%%%%%%%%%%%%%%%%%%%%%%%%%%%%%%%%%%%%%%%%%%%%%%%%%%%%%%%%%%
\subsection{Workload partitioning}\label{subsec:partitioning}

%% Balancing
Hosts and accelerators have different peak (and sustained) performance, so
a careful workload balancing between the two concurrent processors is necessary.
%To do this we find the optimal sizes of the lattice regions to be allocated 
%on the two processors.
%
We model the execution time $T_{\mbox{\tt exe}}$ of our code with the 
following set of equations:
\begin{eqnarray}\label{eq:perf-model}
T_{\mbox{\tt exe}}  &=& \max\{ T_{\mbox{\tt acc}}, T_{\mbox{\tt host}} + T_{\mbox{\tt mpi}} \} + T_{\mbox{\tt swap}}\\
T_{\mbox{\tt acc}}  &=& (LX - 2M)LY \cdot \tau_{d} \\
T_{\mbox{\tt host}} &=& (     2M)LY \cdot \tau_{h} \\
T_{\mbox{\tt mpi}}  &=& \tau_{c} 
\end{eqnarray}
where $T_{\mbox{\tt acc}}$ and $T_{\mbox{\tt host}}$ are the execution 
times of the accelerator and host respectively, $T_{\mbox{\tt swap}}$ 
is the time required to exchange data between host and accelerator at the end of each 
iteration, and $T_{\mbox{\tt mpi}}$ is the time to move data between 
two MPI-processes in a multi-accelerator implementation.
As $T_{\mbox{\tt swap}}$ is independent of $M$,  $T_{\mbox{\tt exe}}$ 
is minimal for a value $M^*$ for which the following equation holds:
\begin{equation}\label{eq:param-tuning}
T_{\mbox{\tt acc}}(M^*) = T_{\mbox{\tt host}}(M^*) + T_{\mbox{\tt mpi}}(M^*)
\end{equation}

Our code has an initial auto-tuning phase, in which it runs a set of mini-benchmarks 
to estimate approximate values for $\tau_{d}, \tau_{h}, \tau_{c}$.
These are then inserted in \autoref{eq:param-tuning} 
to find $M^*$, an estimate of the value of $M$ that minimizes time to solution.

%%%%%%%%%%%%%%%%%%%%%%%%%%%%%%%%%%%%%%%%%%%%%%%%%%%%%%%%%%%%%%%%%%%%%%
\begin{figure}
\centering
\includegraphics[width=.49\textwidth]{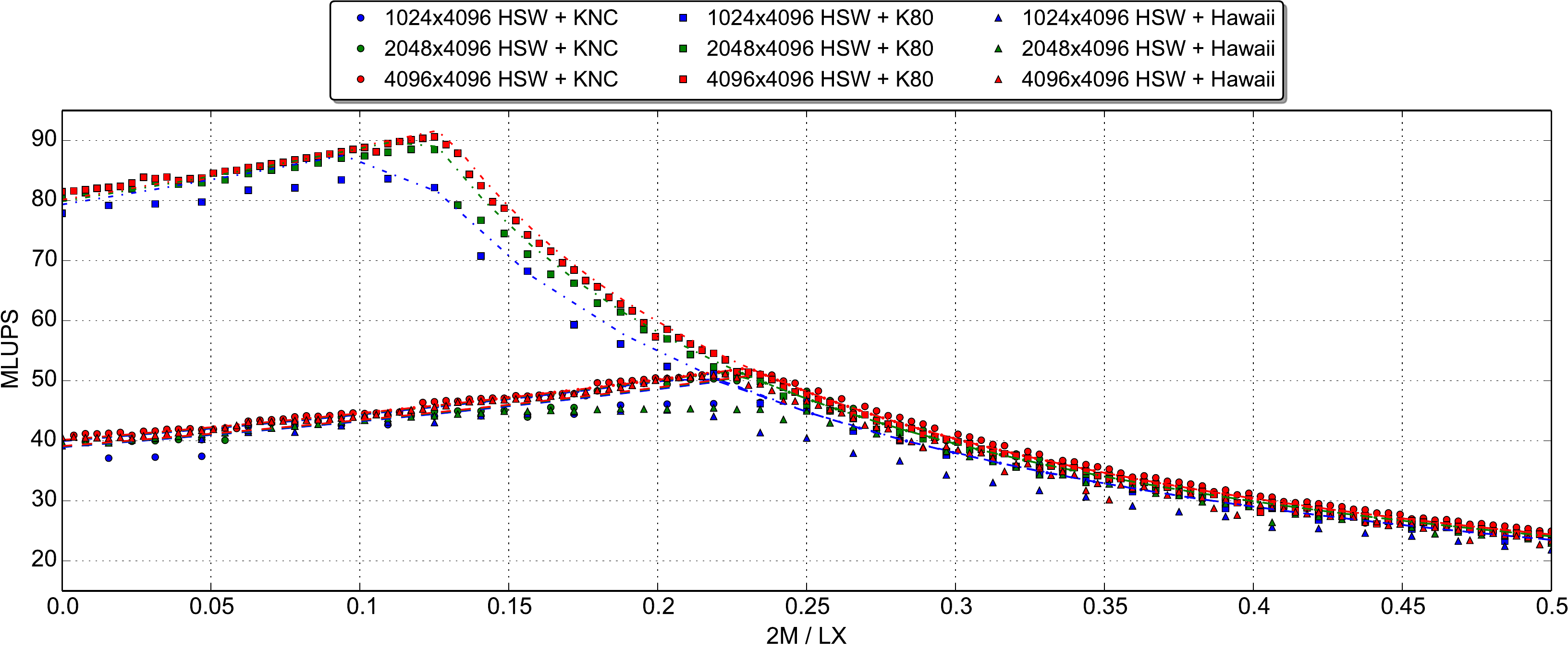}
\caption{ 
Performance of the heterogeneous code (measured in MLUPS, 
see the text for definition) for all three platforms, as a function of the fraction of lattice sites 
($2M/LX$) mapped on the Haswell (HSW) host CPU. KNC is the Intel Knights Corner accelerator, 
K80 is the NVIDIA Tesla GPU and Hawaii is the AMD GPU. 
Dots are measured values, dashed lines are the prediction of our model.  
}
\label{fig:cmp-lbalance-mlups}
\end{figure}
%%%%%%%%%%%%%%%%%%%%%%%%%%%%%%%%%%%%%%%%%%%%%%%%%%%%%%%%%%%%%%%%%%%%%%

In~\autoref{fig:cmp-lbalance-mlups} we show the performance of our 
code for three different lattice sizes as a function of $2M/LX$, the fraction 
of lattice sites that we map on the host CPU. 
We have run our tests on two different machines, the {\em Galileo} HPC system installed 
at {\em CINECA} and the {\em Etna} machine. {\em Galileo}
has two different partitions, one with K80 GPUs 
and one with KNC accelerators. The {\em Etna} machine is 
part of the {\em COKA} experimental cluster at {\em Universit\`a di Ferrara}, and has two 
AMD Hawaii GPUs.
Both machines use as host processor an 8-core Intel Xeon E5-2630v3 CPU 
based on the {\em Haswell} micro-architecture, and each host has one 
attached accelerator. 
Performance is measured using the {\em MiLlion Updates per Second} (MLUPS) 
metric, a common option for this class of applications.
In \autoref{fig:cmp-lbalance-mlups} dots are measured values, while lines are our predictions. 
Our auto-tuning strategy predicts performance with good accuracy,  
and estimates the workload distribution between host and device 
for which the execution time reaches its minimum.
An interesting features of this plot is the fact that the optimal 
point is -- for each accelerator -- a function of $2M/LX$. 
As expected, for values of $M < M^*$ and $M > M^*$ performances decrease  
because the workload is unbalanced either on the accelerator  
or on the host side; results at $2M/LX = 0$ correspond to earlier 
implementations in which critical kernels are fully offloaded to 
accelerators; 
%
% CHECK
we see that running these kernels concurrently on host and accelerators 
(KNC and K80) performances increases approximately by $10-20\%$.
Finally, as $M$ becomes much larger than $M^*$, 
all lines in the plot fall on top of each other, as in this limit 
the host CPU handles the largest part of the overall computation.

%%%%%%%%%%%%%%%%%%%%%%%%%%%%%%%%%%%%%%%%%%%%%%%%%%%%%%%%%%%%%%%%%%%%%%
\begin{figure}
\centering
\includegraphics[width=.49\textwidth]{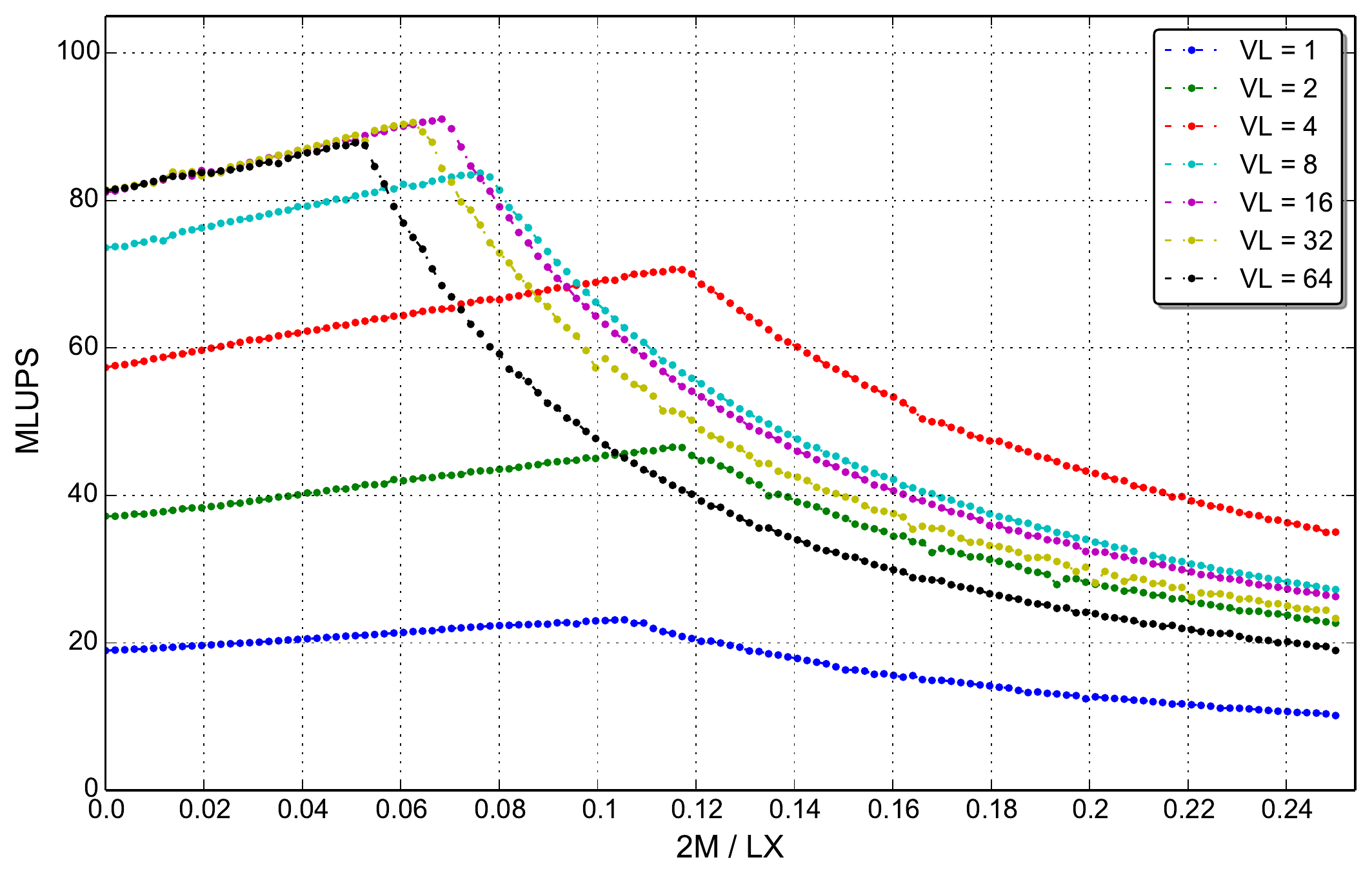}
\caption{
Performance (measured in MLUPS) for different values of the
{\tt VL} parameter in the CaoSoA data layout; results are for a 
Tesla K80 GPU.
}
\label{fig:k80-lbalance-vl-mlups}
\end{figure}
%%%%%%%%%%%%%%%%%%%%%%%%%%%%%%%%%%%%%%%%%%%%%%%%%%%%%%%%%%%%%%%%%%%%%%

%%%%%%%%%%%%%%%%%%%%%%%%%%%%%%%%%%%%%%%%%%%%%%%%%%%%%%%%%%%%%%%%%%%%%%
\subsection{Fine-tuning of data-layout cluster size}

An important parameter of the CAoSoA layout is the 
cluster size {\tt VL}, as performance depends significantly 
on its value. This parameter, whose optimal value correlates with the 
hardware features of the target processor, affects data allocation in 
memory and must be fixed at compile time. 

\begin{figure}
\centering
\includegraphics[width=.49\textwidth]{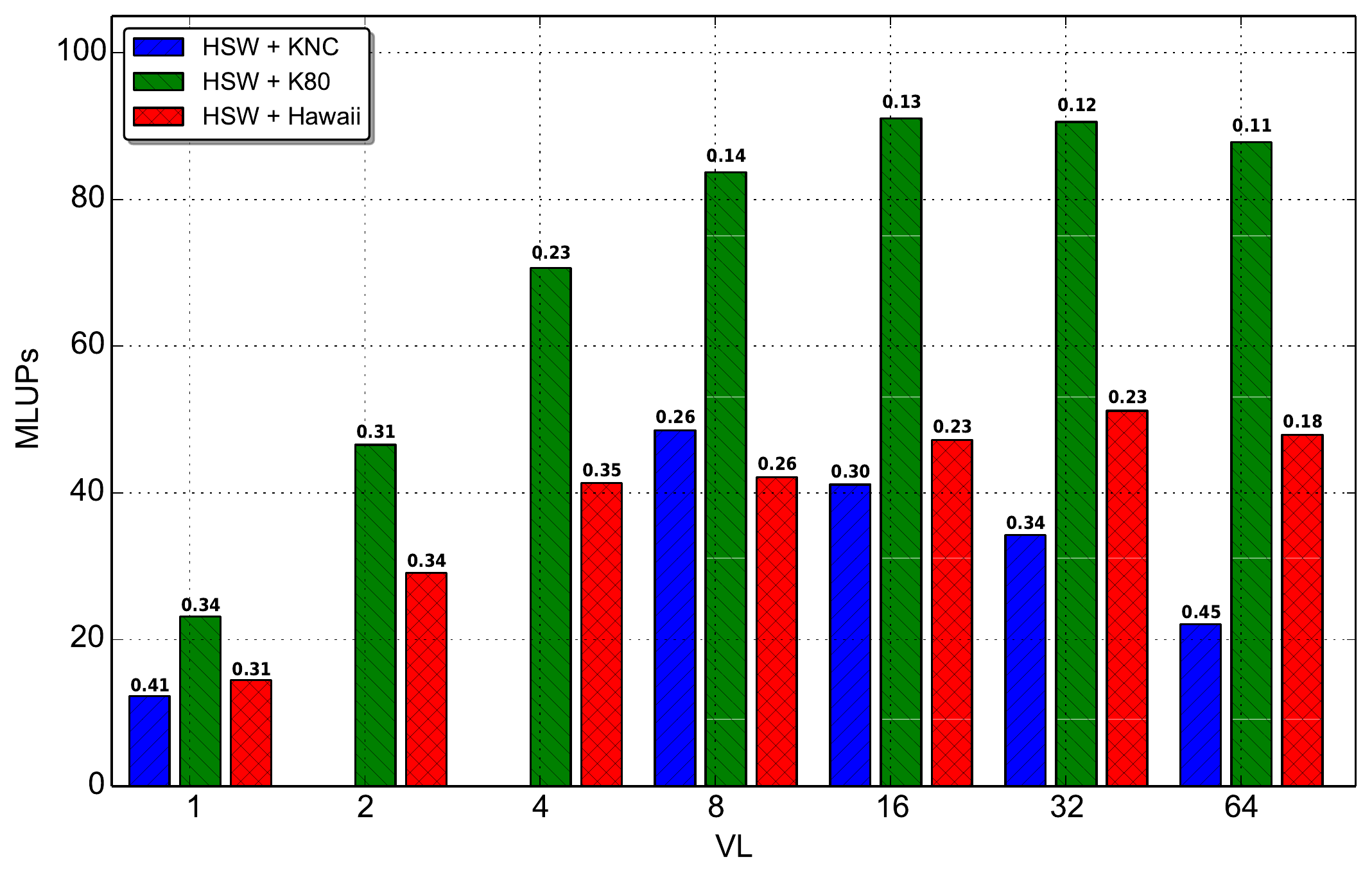}
\caption{ 
Impact on performance of different cluster sizes (VL) in the
CaoSoA data-layout for several accelerator choices. 
}\label{fig:cmp-vl-mlups}
\end{figure}

In~\autoref{fig:k80-lbalance-vl-mlups} we show the impact on performances 
(measured again in MLUPS) of our code running on a node with K80 GPUs and 
using different choices for {\tt VL}. 
We see that, for this processor, a wrong choice may reduce performance 
by large factors ($\simeq 5$); good news is that there is a reasonably 
large interval {\tt VL=16,32,64} for which performance is close 
to its largest value.
We have done the same measurements for all kind of nodes available with KNC and 
AMD GPUs, and then picked, for each node and each value of 
{\tt VL}, the best operating point in terms of $2M/LX$. 
In this way, we select the highest performance for each combination of host 
and accelerator. 
These results are collected in~\autoref{fig:cmp-vl-mlups} as a 
function of {\tt VL}; on top of each bar we report the corresponding value 
of $2M/LX$. 
We see that GPUs are more robust than the KNC against a non optimal 
choice of {\tt VL}: for the former processors performance remains 
stable as long as {\tt VL} is large enough, while for the latter 
only one or two {\tt VL} values allow to reach the highest performance. 
Fortunately enough, there is a window of {\tt VL} values for which all 
systems are close to their best performance. 
%In summary, for this problem, 
%the combination of a Haswell host and a K80 GPU operating at the optimal point 
%is $2 \times$ faster than all other options. \red{toglierei quest'ultimo commento,
%non mi sembra un "summary" della sezione}

%%%%%%%%%%%%%%%%%%%%%%%%%%%%%%%%%%%%%%%%%%%%%%%%%%%%%%%%%%%%%%%%%%%%%%
%%%%%%%%%%%%%%%%%%%%%%%%%%%%%%%%%%%%%%%%%%%%%%%%%%%%%%%%%%%%%%%%%%%%%%

%%%%%%%%%%%%%%%%%%%%%%%%%%%%%%%%%%%%%%%%%%%%%%%%%%%%%%%%%%%%%%%%%%%%%%
\subsection{Performance prediction on new hardware}\label{subsec:prediction}

A further interesting result of the model developed in \autoref{subsec:partitioning}, 
is that we can use it to predict to which extent the performance of our codes 
is affected if either the host CPU or the accelerator is replaced by a different processor; 
in particular one may ask what happens if announced but not yet available processors or 
accelerators are adopted. 
One such exercise replaces the host processor that we have used for 
our previous tests with the new Intel multicore Xeon E5-2697v4, 
based on the latest {\em Broadwell} micro-architecture.
We have run our code on a {\em Broadwell} processor with no attached 
accelerators and measured the host-related performance parameters 
used in \autoref{eq:perf-model}; 
we have then used these parameters to estimate the expected performance 
of a would-be machine whose nodes combine Broadwell hosts with either 
K80 or KNC accelerators.
%
%%%%%%%%%%%%%%%%%%%%%%%%%%%%%%%%%%%%%%%%%%%%%%%%%%%%%%%%%%%%%%%%%%%%%%
\begin{figure}
\centering
\includegraphics[width=.49\textwidth]{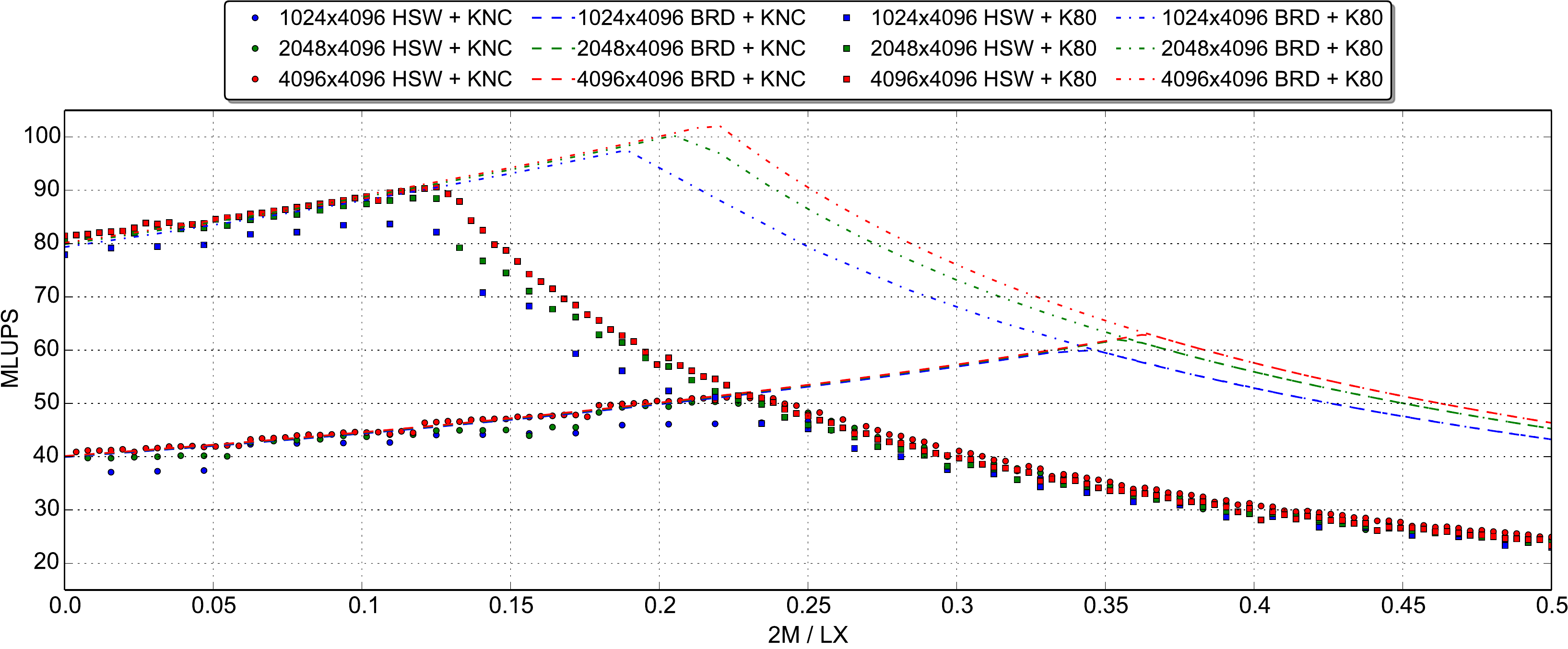}
\caption{ 
Performance predictions (dashed-lines) of our model for a would-be 
system using as host the recently-released Broadwell (BRD) CPU compared 
with measured data on a Haswell (HSW) CPU (dots). Measurements refer to three 
different lattice sizes.
}\label{fig:cmp-lbalance-mlups-broadwell}
\end{figure}
%%%%%%%%%%%%%%%%%%%%%%%%%%%%%%%%%%%%%%%%%%%%%%%%%%%%%%%%%%%%%%%%%%%%%%%
%
Results are shown in \autoref{fig:cmp-lbalance-mlups-broadwell} 
where we compares the measured performance on the present hardware (dots) 
with the predictions of our model (dashed lines).  
%
%%% CHECK
We see that using this new more powerful processor, performance for our code
would improve by approximately $20\%$, when perfectly balancing the workload
between host and accelerator. 
As expected the  improvement is the same for both types of accelerators, since we
have only virtually replaced the host processor.
Another interesting feature of the plot is that the model predictions overlap
with measured data when $2M/LX$ values tends to zero; this is expected, since in
this case the fraction of lattice sites mapped on the host-CPU tends to zero and
the execution time is dominated by the accelerator.
A similar analysis might be performed, for instance, to assess the overall 
performance gains to be expected when next generation GPUs become available.

%%%%%%%%%%%%%%%%%%%%%%%%%%%%%%%%%%%%%%%%%%%%%%%%%%%%%%%%%%%%%%%%%%%%%%
\begin{figure*}
\centering
\includegraphics[width=.99\textwidth]{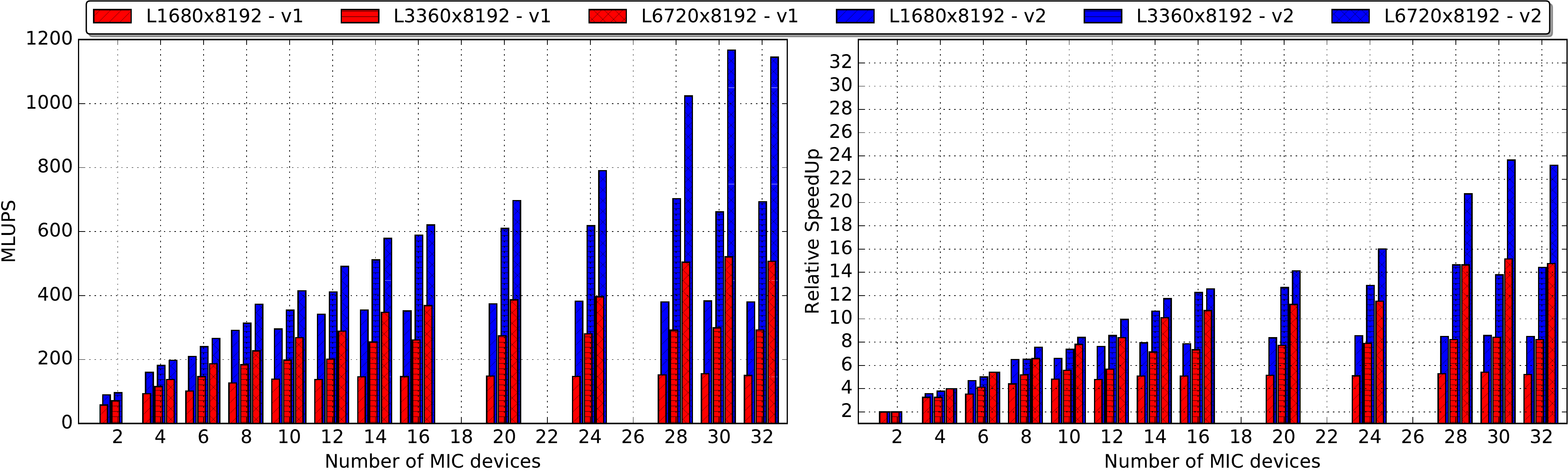}
\includegraphics[width=.99\textwidth]{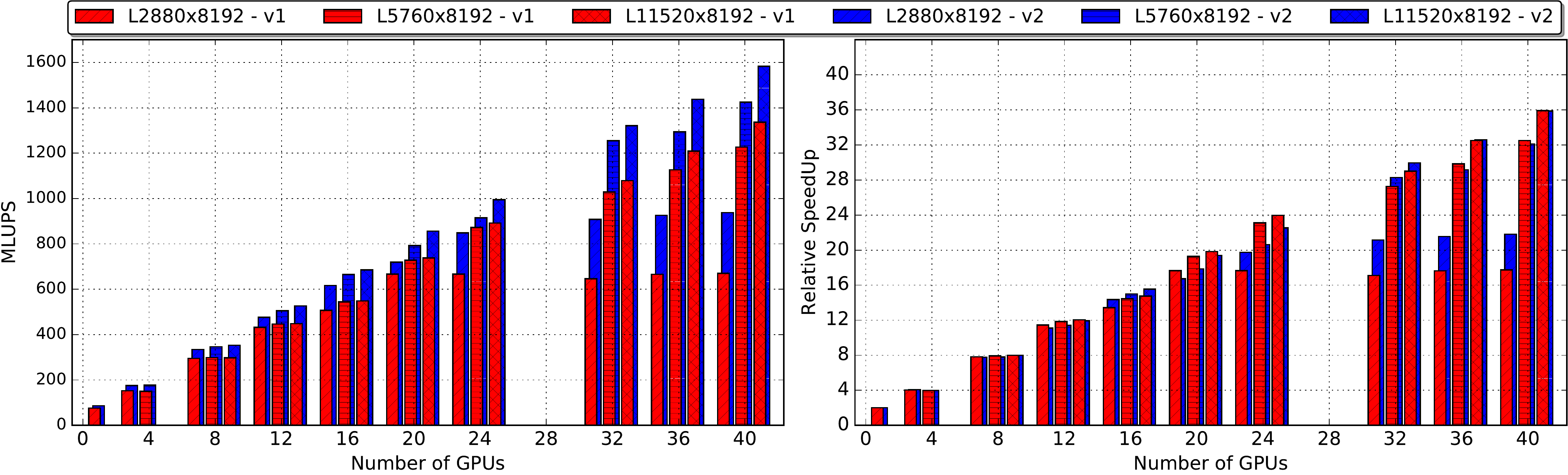}
\caption{
Multi-node scalability results measured on the KNC (top) and K80 (down) 
partitions of the Galileo cluster. 
We compare performances -- MLUPS and relative speedup -- 
on three different lattice sizes of the heterogeneous code described in this paper (v2) with
an earlier version (v1) with all critical computational kernels running on accelerators only.
}\label{fig:knc-cmp-scale}\label{fig:galileo-k80-cmp-scale}
\end{figure*}
%%%%%%%%%%%%%%%%%%%%%%%%%%%%%%%%%%%%%%%%%%%%%%%%%%%%%%%%%%%%%%%%%%%%%%

%%%%%%%%%%%%%%%%%%%%%%%%%%%%%%%%%%%%%%%%%%%%%%%%%%%%%%%%%%%%%%%%%%%%%%
\section{Scalability performances}\label{sec:scalability}

In this section we analyze scalability performances of our codes running 
on the {\em Galileo} machine, both on the K80 and on the KNC partition. 

In \autoref{fig:knc-cmp-scale} top-left we show the performance of
our code running on larger and larger KNC partitions of the {\em Galileo} cluster,
and for several physically relevant sizes of the physical lattice, showing 
the scaling results of this code. We compare with a previous {\tt v1}
implementation  of the code running all kernels on the KNC accelerator.
\autoref{fig:knc-cmp-scale} top-right shows the same data as the previous picture
showing however the speed-up factor as a function of the number of accelerators.
One easily sees that the new heterogeneous version of the code is not only faster 
than its accelerator-only counterpart but also has a remarkably better behavior 
from the point of view of hard scaling. This is due to the fact that 
data moved through MPI communications between different processing nodes is always 
resident on the host, saving time to move them to and from KNC accelerator.
All in all, for massively parallel runs on many accelerators, the heterogeneous 
code extracts from the same KNC-based hardware system roughly $2 \times$ 
larger performance than the {\tt v1} version. 

In \autoref{fig:galileo-k80-cmp-scale} bottom-left we show the performance of
our code running on the K80 partition. In this case, due to the larger 
difference in performance between the host-CPU and the accelerator we have a 
different behavior. 
Version {\tt v2} is faster than version {\tt v1}, but the gain is less compared to KNC 
because the gap in performance between the K80 GPU and the host CPU is larger.  
Scalability, see~\autoref{fig:galileo-k80-cmp-scale} bottom-right, is good as version {\sf v1} 
because in both implementations MPI communications a fully overlapped with computation (\cite{hpcs15}).

%%%%%%%%%%%%%%%%%%%%%%%%%%%%%%%%%%%%%%%%%%%%%%%%%%%%%%%%%%%%%%%%%%%%%%%%
\section{Analysis of results and conclusions}\label{sec:conclusion}

In this section we analyze our results and outline our conclusions.

The first important contribution of this work highlights the critical 
role played by data layouts in the development of a common LB code  
that runs concurrently on CPUs and accelerators and is also {\em performance-}portable 
onto different accelerator architectures.
The crucial finding in this respect is that data memory organization should 
support at the same time efficient memory accesses and vector processing. 
This challenge is made more difficult because different kernels 
(in our case the critical {\sf propagate} and {\sf collide} routines) 
have conflicting requirements. 
\autoref{tab:propagate-kernel} substantiates the relevance of this problem 
and quantifies the improvements that we have achieved. 
Previously used data structures are the AoS layout supporting data-locality, 
and the {\em SoA layout} already known to exploit 
more vectorization especially for GPUs. 
The problem with these two layouts is that the former allows better 
performances for Intel architectures but is dramatically bad for GPUs, 
with performance losses that are up to $10X$ for {\sf propagate} and 
$2X-5X$ for {\sf collide}.
The latter scheme is very efficient on GPUs but it fails on CPUs, 
as it limits code vectorization and causes memory management 
overheads (like TLB and cache misses). 

This paper introduces two slightly more complex data layouts, 
the CSoA and the CAoSoA to exploit vectorization on both classes 
of architecture and still guaranteeing efficient data memory accesses and 
vector processing for both critical kernels. 
These data-structures differ from those used in \cite{shet2} and \cite{pedro15} 
in the way populations are packed into SIMD vectors; this allows  
to properly translate operations involved in the {\tt propagate} kernel into 
SIMD instructions, and to perform aligned memory accesses. 

The CSoA layout improves performance on Intel architectures by factors 
$2X$ over the AoS for {\sf propagate} and $1.5X$ for {\sf collide}. 
On GPUs it has also a marginal improvements compared to the already 
very good results that GPUs have with the original SoA layout.
A final improvements is given by the CAoSoA layout, that further 
increases data locality without introducing vectorization penalties. 
Again, performance remains substantially stable on GPUs in this case, 
while there is still a further improvements on Intel architectures for 
{\sf collide} ($\approx 10-20\%$) and a corresponding penalty for 
{\sf propagate}; since the former routine has a bigger impact on overall 
performance, the CAoSoA layout is the most efficient to be used 
for the whole code. 

The definition of the appropriate layouts has allowed to code our LB 
application in a common program that executes concurrently on host CPUs and 
all kinds of accelerators, obviously improving the portability and 
long-term support of this code. 

Another important contribution of this paper is the development of 
an analytic model (see \autoref{subsec:partitioning}) able to predict the optimal 
partitioning of the workload among host-CPU and accelerators; using this 
model we can automatically tune this parameter for best performance on 
running systems, or predict performances for not yet available hardware 
configurations (see \autoref{subsec:prediction}). 

% CHECK
The final result of this contribution is that single 
node performances can be improved by $\approx 10-20\%$ w.r.t. earlier 
implementations that simply wasted the computational power offered 
by the host CPU. 
Our implementation also significantly improves the (hard)-scaling behavior on 
relatively large clusters, see \autoref{fig:knc-cmp-scale}. 
This follows from the fact that node-to-node communications in our 
code do not imply host-accelerator transfers. 
This improvement is very large on KNC-accelerated clusters, while on GPUs,
due to the larger performance unbalance between host and accelerator, 
the observed improvement is smaller.

% commenti finali
In conclusion, an important result of this work is that it is possible 
to design and implement directive-based codes that are performance-portable 
across different present -- and hopefully future -- architectures. 
This requires an appropriate choice of the data layout, able to meet 
conflicting requirements of different parts of the code and to match 
hardware features of different architectures. 
Defining these data layout is today in the hands of programmers, and  still out
of the scope of currently available and stable compilers  commonly used in the
HPC context.
For this reason, on a longer time horizon, we look forward to further 
progress in compilers allowing data definitions that abstract 
from the actual in-memory implementation, and of tools able to make 
appropriate allocation choices for specific target architectures.

%%%%%%%%%%%%%%%%%%%%%%%%%%%%%%%%%%%%%%%%%%%%%%%%%%%%%%%%%%%%%%%%%%%%%%%%

\begin{acks}
This work was done in the framework of the COKA, COSA and Suma projects of INFN. 
We thank CINECA (Bologna, Italy) for access to the Galileo computing cluster.
AG has been supported by the European Union's Horizon 2020 research and
innovation programme under the Marie Sklodowska-Curie grant agreement No 642069.
\end{acks}

%%%%%%%%%%%%%%%%%%%%%%%%%%%%%%%%%%%%%%%%%%%%%%%%%%%%%%%%%%%%%%%%%%%%%%%%

\bibliographystyle{SageH}

%%%%%%%%%%%%%%%%%%%%%%%%%%%%%%%%%%%%%%%%%%%%%%%%%%%%%%%%%%%%%%%%%%%%%%%%

\end{document}